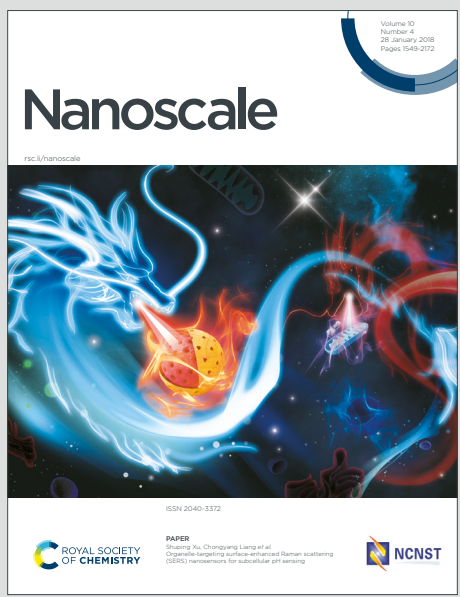

# Unveiling the potential of BCN-Biphenylene monolayer as a high-performance anode material for alkali metal ion batteries: A first-principles study

Ajay Kumar[a] and Prakash Parida[a]*

[a] Department of Physics, Indian Institute of Technology Patna, Bihta, Bihar, India, 801106

*Corresponding author: - pparida@iitp.ac.in

*Abstract*

Inspired by the freshly synthesized two-dimensional biphenylene carbon network, which features a captivating combination of hexagonal, square, and octagonal rings, we explore a similar biphenylene network composed of boron, carbon, and nitrogen (bpn-BCN) using first-principles calculations. There are six possible phases of borocarbonitrides, which are isoelectronic to biphenylene carbon networks with a stoichiometric ratio of 1:1:1 for boron (B), carbon (C), and (N) atoms. All possible isoelectronic structures of the BCN combination of biphenylene networks are found to be stable, according to first principles calculations. However, the geometry has a relatively large number of robust C-C, B-N bonds and strong partial ionic-covalent B-C and C-N bonds inside these bpn-BCN monolayers is effectively more stable. Further, we employed first-principles calculations to investigate the electrochemical properties of the most stable geometry of BCN biphenylene as a potential anode material for alkali metal (AM) ion batteries. A global search has been made to find the most favourable alkali metal ion adsorption sites. The biphenylene monolayer has octagonal, square, and hexagonal motifs with different adsorption strengths. Further, the partial ionic bond of B-N (due to the electronegativity difference) also supports the metal alkali ions for adsorption. The electronic properties of the stable phase of bpn-BCN reveal its narrow bandgap semiconductor nature. The ion diffusion calculations show a low activation barrier for Li, Na, and K of 0.65 eV, 0.26 eV, and 0.23 eV, respectively, indicating a fast charge/discharge rate. Furthermore, the theoretical capacities of the BCN biphenylene monolayer for Li (1057.33 mAhg$^{-1}$), Na (647.27 mAhg$^{-1}$), and K (465.98 mAhg$^{-1}$) are found to be greater than those of commercial graphite. The average open-circuit voltage for AM decreases with increased metal ion concentrations. It falls within a reasonable range of 0.34–1.89 V. Our results show that BCN biphenylene monolayer could be a promising anode material in alkali metal ion rechargeable batteries.

## 1. Introduction

With the rising demand for large-scale energy storage devices such as electric vehicles and power backups, a major concern is the high cost of a rechargeable battery system.[1,2] There is an urgent need to focus on low-cost and high-storage electrodes for batteries. Rechargeable lithium-ion batteries (LIBs) with high specific capacity, high power density, and extended cycle life are in high demand to enable wider adoption of electric vehicle technology.[3,4] Lead-acid and lithium-ion batteries, the most often used batteries, are very effective but have some limitations in production costs, safety, and durability.[5-8] Other alkali metal ions, such as those used in sodium (Na) and potassium (K) ion batteries, have recently gained significant research interest due to their abundant availability, cost-effective extraction, and large storage capacities.[9,10] Since Li, Na, and K belong to the same alkali metal group, conducting comparative research on alkali metal ion batteries (AIBs) is advisable.

The performance of Li and other metal ion batteries, particularly, depends on the qualities of the constituent electrode. Graphite is a commercially used anode material due to its good cyclic endurance, high coulombic efficiency, and inexpensive cost.[11] However, the comparatively low theoretical specific capacity (372 mAhg$^{-1}$) prevents it from being further used. Other well-studied anode materials, such as transition metal oxides[12] and silicon[13], have important concerns, such as large irreversible capacity, low lithium diffusivities, and large volume change. As a result, there is an urgent need to look for new anode materials to boost further the performance of Li and other metal ion batteries.

2D materials offer significant advantages over bulk systems due to their unique structural and electronic properties, especially their high surface-to-volume ratio, which enhances their interaction with foreign atoms (adsorbates) during adsorption.[14] This increased surface area provides more active sites, beneficial for energy storage, and their high surface energy promotes metal atom adsorption, enhancing storage capacity.[15, 16] Additionally, 2D materials allow freer movement of metal ions, leading to a higher diffusion coefficient compared to bulk crystals, which is crucial for battery performance. While bulk crystals generally exhibit greater structural stability, the stability of 2D materials should be optimized and comparable to the bulk system to ensure reliable and long-lasting battery performance through many charging and discharging cycles.[17-19] Graphene and other 2D crystals and their hybrid structures can be studied as attractive options in light of the growing demand for more efficient rechargeable energy storage devices due to their enormous surface area, excellent mechanical flexibility, and high electron mobility.[20-22] Recent findings indicate that 2D materials exhibit higher specific capacities compared to commercially available graphite. For instance, graphene has a specific capacity of 540 mAhg$^{-1}$ for Li,[23] $MoS_2$ has 389 mAhg$^{-1}$ for Na,[24] silicene boasts 954 mAhg$^{-1}$ for Li,[25] and porous borophene reaches 1157 mAhg$^{-1}$ for Li.[26] Despite certain advancements in these electrode materials, there is still a need for novel 2D electrode materials. Therefore, it is necessary to search for more two-dimensional anode materials in order to improve the electrochemical characteristics of alkali metal ion batteries.

In recent years, a wide range of two-dimensional (2D) metal-free materials has been studied, including mono-element boron, carbon and phosphorous, binary boron nitride, carbon nitride and boron carbide, and ternary boron-carbon-nitrogen combinations.[26-32] Both experimental and theoretical findings result in the potential of producing stable 2D materials from earth-abundant and metal-free elements, which opens up new paths for photochemical, electrochemical, and electrical technologies.[33-35] The ternary graphene-like honeycomb structure of boron, carbon and nitrogen (g-BCN) with various stoichiometric ratios has been successfully synthesized and used as an efficient catalyst in the electrochemical and photochemical processes.[36-38] Li et al. produced h-BN decoration on graphene sheets by controlling the doping sequence of boron-nitrogen components, which was employed in photocatalytic hydrogen generation.[37] S. Wang et al. reported the electrocatalytic activity even better than the Pt-based electrocatalyst. The metal-free g-BCN electrocatalyst with different B/N co-doping levels has been synthesized by thermal annealing of graphene oxides in the presence of boric acid and ammonia.[38] Shucheng et al. synthesized a range of carbon materials incorporating isolated h-BN patches within the carbon framework, showcasing their potential as catalysts for $H_2O_2$ production. Through modulation of initial co-monomer precursor ratios, they effectively managed Brunauer-Emmett-Teller (BET) surface area and the overall content of B and N dopants within the carbon structures.[39] Sumit et al. synthesized 2D graphene-like BCN using bis-BN cyclohexane ($B_2N_2C_2H_{12}$) as a precursor molecule. The process involved thermally induced dehydrogenation to form an epitaxial monolayer on Ir(111) via covalent bonding. First-principles calculations predict that the resulting material has a direct electronic band gap intermediate between the gapless nature of graphene and the insulating properties of h-BN.[40] In addition to these experimental studies, computational research were also done to study the thermal, electrical, and mechanical stability of ternary BCN monolayers with various chemical compositions and to identify potential theoretical applications for these materials.[41-44]. For illustration, Thomas et al. conducted a comparative study of h-BCN monolayers with graphene and h-BN using classical molecular dynamics simulations. They found that the mechanical strength of h-BCN is highly anisotropic and temperature-dependent. While h-BCN is mechanically less stable than graphene, it has better mechanical stability compared to h-BN.[45] Besides the success of both experimental and theoretical studies on 2D graphene-BCN compounds, is it conceivable to investigate a structure other than the honeycomb-like network (graphene) for BCN compositions? Theoretically, several carbon allotropes have been investigated, some of which are experimentally synthesized. For example, non-graphitic carbon networks include graphenylene, a special porous network of nondelocalized carbon atoms,[46] 2D networks of the carbon allotropes T-carbon,[47] and graphyne were also studied.[48]

To find the non-graphitic BCN, which contains a uniform distribution of B/C/N atoms and distinctive chemical connections, the biphenylene-carbon atom network (BPN) has been considered. Fan et al. synthesized an ultra-flat biphenylene carbon sheet on a gold surface from $sp^2$-hybridized carbon atoms that formed four-, six-, and eight-membered motifs.[49] Meanwhile, BPN is a well-studied material for theoretically and experimentally investigating numerous applications.[50-53] The metallic nature of pristine BPN has been reported as an anode material for sodium-ion batteries.[54] The B-N composition of the biphenylene network has been explored in terms of electronic and thermoelectric properties.[55] In the meantime, the B-N composition of biphenylene monolayer is a large bandgap semiconductor, which is not a good choice for anode material. Further, due to the similar atomic sizes of B, C, and N, 2D B- and N-doped graphene structures have attracted much interest as prospective anode materials for LIBs and NIBs.[56, 57] The primary advantage of employing the BCN structure as an anode material lies in the partially ionic bond between B and N. This bond creates an electron deficiency on B and an extra charge on N, facilitating the adsorption of alkali atoms. These promising developments encourage us to contemplate using B and N-doped biphenylene as an anode material for alkali-ion batteries. Therefore, the hybrid bpn-BCN monolayer will be a better choice for exploring the anode material for alkali ion batteries.

In this study, six isoelectronic phases of boron, carbon, and nitrogen atoms, with an equal stoichiometric ratio (1:1:1) per unit cell, has been explored within the biphenylene network (bpn-BCN), mirroring the experimentally synthesized carbon biphenylene monolayer. Choosing borocarbonitrides over pristine carbon offers a distinct advantage—partial ionic bonds enhance the intense adsorption of metal atoms. However, the only B-N biphenylene network has a wider bandgap, unfavourable for anode materials. Materials with narrow bandgap or metallic characteristics are preferred because of good ionic and electronic conductivity. Typically, semiconductors characterized by a narrow bandgap transition into a conductive state upon metal atoms adsorption, making them a favourable choice for anode materials. We examine the structural and electronic properties of the bpn-BCN biphenylene monolayer and check its substantiality for anode material in alkali metal ion batteries via first principle calculations. Our study primarily aims to investigate the structural stability, electronic characteristics, and various electrochemical properties such as adsorption behaviour, diffusion kinetics, charge storage capabilities, and open circuit voltages of bpn-BCN. These features are crucial in determining the suitability of BCN biphenylene material as an anode material for alkali-ion batteries.

## 2. Computational Details

We have examined the multiple phases of a 2D biphenylene-like BCN network within the density functional theory framework through the Vienna ab initio simulation package (VASP 6.1.0).[58] All the calculations have been done using the exchange-correlation function of the generalized gradient approximation Perdew-Burke-Ernzerhof (GGA-PBE).[59] A vacuum layer of 20 Å along the z-direction has been used in order to maintain the negligible effect of interlayer interactions. The geometry has been fully optimized until the force acting on each atom is less than 0.001 eV/Å, and total energy convergence for self-consistent calculation is tolerance upto $10^{-10}$ eV. The plane wave energy cutoff has been set to 600 eV throughout the calculations. The k-grid with a mesh size of 15×12×1 in the first Brillouin zone has been used. The phonon dispersion spectra of different phases of BCN have been calculated in the framework of the phonopy package using the finite displacement method.[60] The supercell of dimensions 4×3×1 has been used for each phase of BCN. Furthermore, force constants have been calculated using a VASP-phonopy interface. The ab initio molecular dynamics (AIMD) simulations have been carried out in the NVT ensemble at various temperatures with a time step of 1 *fs* over a 1-10 *ps* range.[61] A supercell of 4×3×1 and k-grid of 9×7×1 has been used in each phase of the BCN hybrid system. A Nose-Hoover thermostat has been used to maintain the temperature values of 300 K and 800 K. Further, the convergence charge density has been used to calculate the electron localization function (ELF), non-self-consistent electronic band structure, and projected density of states. The electronic band structures of different BCN phases have been plotted along a high symmetry path using GGA-PBE exchange-correlational functional. The electronic band structure has also been calculated in the Heyd–Scuseria–Ernzerhof (HSE) hybrid functional framework to ensure the robustness of electronic properties.[62] The van der Waals (vdW) interactions between the adatoms with monolayer have been considered using the Grimme zero damping DFT-D3 approach for electrochemical properties.[63] To quantify the interaction between alkali atoms and the nearest atom of bpn-BCN after adsorption, we used the LOBSTER program to calculate Crystal Orbital Hamiltonian Population (COHP) and Integrated COHP (ICOHP).[64] The climbing image-nudged elastic band[65] (CI-NEB) approach has been used to find the transition state (TS) structure and diffusion barrier of a metal-adsorbed bpn-BCN monolayer. The thermal stability of pristine and metal-adsorbed bpn-BCN monolayers has been investigated using AIMD simulations in the framework of the NVT ensemble. A supercell of 3 × 3 × 1 has been generated for AIMD at 500K with a time step of 1*fs* over a range of 0-5 *ps*. We have used visualization for electronic structural analysis (VESTA) software to visualize the atomic structures and charge densities.[66]

## 3. Results and Discussions
### 3.1. Optimized crystal geometry of bpn-BCN monolayers

The optimised atomic structure of all possible six configurations of bpn-BCN compositions has been depicted in **Fig. 1**. The biphenylene-like BCN crystal structures show octagon-hexagon-squares motifs with an oblique ($a \neq b; a \angle b \neq 90°$) primitive unit. The bpn-BCN is composed of six atoms with equal stoichiometric ratios of B(green), C(brown), and N(silver), and the primitive unit cell is shown as a black box in **Fig. 1**. Let's name these six phases as α-BCN, β-BCN, γ-BCN, δ-BCN, η-BCN, and θ-BCN, which are further marked in **Fig. 1(a-f).** It is to be observed that there is cis- and trans-like bonding of atoms about the central bond in a unit cell. For illustration, α-BCN and δ-BCN have the same B-B bond at the centre of a unit cell. The difference is that the former has cis- and the latter has trans-coordination with C and N about the B-B bond. Similarly, the pair of β- and η-BCN, as well as γ- and θ-BCN, both share the cis-trans atomic arrangement with a centre bond of C-C and N-N, respectively. The optimized geometry of α-, β-, γ-BCN (all cis-) phases have a rectangular 2D crystal system with symmetry space group *P1* whereas δ-, η-BCN have oblique with space group *P2/m* and θ-BCN has *Pm*. These different symmetries in the crystal structure of bpn-BCN are because of the slight distortion of the rectangular unit cell to oblique. α-, β-, γ-BCN have a *P1* symmetry group and have rectangular primitive unit cells (a=b, χ=90°, where χ is the angle between a and b). Similarly, δ-, η- and θ-BCN (all trans-) have oblique unit cells where the angle for δ-, η-BCN phases is obtuse (a=b, χ>90°) and for θ-BCN is acute (a=b, χ<90°). This effect will be seen in the high symmetry path chosen in the 1st Brillouin zone for dispersion plots. The lattice parameters, bond types, bond length and their counts per unit cell have been reported in **Table S1** in the supplementary information (SI).

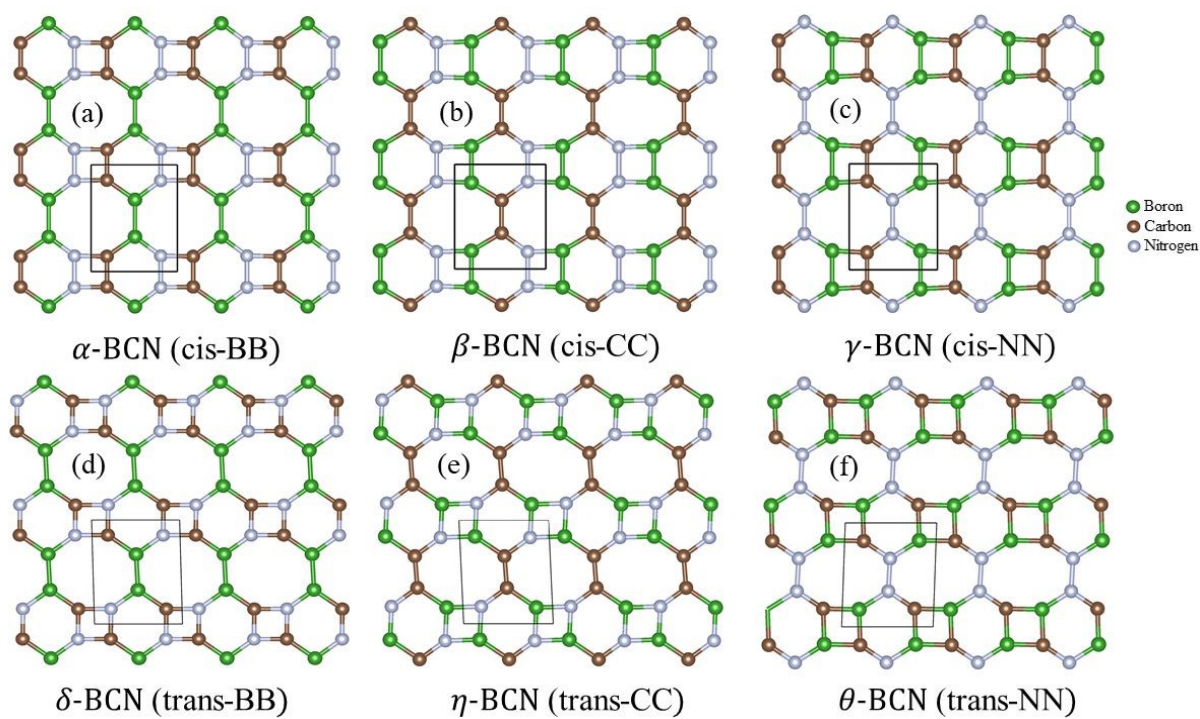

*Fig. 1* *Optimized atomic structures of different phases of the bpn-BCN monolayers. The brown, grey and green spheres refer to C, N, and B atoms.*

In a biphenylene network, each unit cell contains nine necessary linkages, as shown in **Fig. S1** of the supplementary information (SI). All possible bonds in the BPN monolayer are C-C. However, in hybrid bpn-BCN phases, there are six different types of bonds: B-B, B-N, B-C, C-C, C-N, and N-N. These different phases of BCN are due to different bonds and their location in the unit cell. For example, in α-BCN and δ-BCN, B-B shares an octagon with a distinct environment, and in β-BCN and γ-BCN share a hexagon and a square ring with different atoms around. These bonds influence the electronic properties of distinct phases, which will be studied more below.

### 3.2. Structural stability
#### 3.2.1. formation and cohesive energy

To explore the thermodynamic stability of the bpn-BCN phases, the formation and cohesive energy have been calculated.[67, 68] Both terms are commonly used to check the stability, with the difference between these terms lying in the utilization of the energy of constituent atoms. Formation energy involves the chemical potential (μ), while cohesive energy utilizes the energy of isolated single atoms. The chemical potential of boron ($μ_B$), carbon ($μ_C$), and nitrogen ($μ_N$) are the energies per atom in their respective stable phases. Similarly, cohesive energy also helps to confirm the thermodynamic stability of the system in terms of isolated single-atom energy. The formation energy ($E_{p-BCN}^{form}$) and cohesive energy ($E_{p-BCN}^{coh}$) per atom has been calculated using the following equations

$$E_{p-BCN}^{coh} = \frac{E_{p-BCN} - (2E_B + 2E_C + 2E_N)}{6} \qquad (1)$$

and,

$$E_{p-BCN}^{form} = \frac{E_{p-BCN} - (2μ_B + 2μ_C + 2μ_N)}{6} \qquad (2)$$

here $E_{p-BCN}$ is the total energy of the $p$ (= α, β, γ, δ, η and θ)-phase of the bpn-BCN monolayer and the energies of isolated single B, C, and N atoms, calculated by placing each atom in a cubic box with a box length of 20 Å. On the other hand, $μ_C$ represents the energy per carbon atom derived from graphite, a stable carbon allotrope. In nitrogen-rich surroundings, the energy per atom ($μ_N$) is determined from the alpha-N2 phase of solid nitrogen, whereas in boron-rich environments, $μ_B$ is calculated from the metallic alpha-B phase. All these values are reported in the SI and have been verified in our previous study.[55] The negative cohesive energy for these monolayers indicates that the free-standing 2D sheet is likely to be stable. On the other hand, formation energy is relative, it may be positive or negative.

To recognize the chemical bonding and stability mechanisms inside these bpn-BCN structures, the iso-surface of electron localized function (ELF) for several phases has been calculated. The ELF is crucial for describing electron localization and chemical bonding in the solids system. For visualization of electron density at a different location, the ELF plots of bpn-BCN phases with an iso-surface value of 0.4 atomic units have been displayed in **Fig. S2**. Moreover, ELF values of 1.0 and 0.5 represent fully localized and partially delocalized electrons, respectively, whereas zero denotes low charge density. In other words, in the ELF plots, red represents the electron deficiency region, and magenta indicates the electron-abundant region. To evaluate if they follow the same mechanism as in the g-BCN monolayer, it is important to carefully study the electronic configurations of the B, C and N in the octagon square and hexagon ring in different phases. The ELF plots indicate that the charge is predominantly concentrated between the bonded atoms throughout the entire B-C-N biphenylene network, highlighting the strong interactions among boron, carbon, and nitrogen atoms. Chemical bonds between identical atoms in these bpn-BCN phases, like B-B, C-C, and N-N, are covalent in nature. On the other hand, partial ionic-covalent interactions would be predicted due to the relative differences in electronegativity among B, C, and N atoms. This electronegativity margin causes an asymmetric distribution of ELF around the N atoms, resulting in the partial ionic-covalent nature of the B-N and C-N bonds.

The stability of these isoelectronic compositions of BCN phases in BPN geometries should be beneficial for explaining different types of strong covalent and ionic-covalent interactions. However, the number of C-C and B-N strong bonds affects the kinetic stability of the phase, further revealing the comparative less cohesive energy. In **eq. 2**, a $E_{p-BCN}^{form}$ indicates the stability of the bpn-BCN system in comparison to stable phases of constituent atoms, while a positive value implies a less of stability. The cohesive energies of p-BCN monolayers reflect the similar sequence of stability of η-BCN (-2.31 eV) < δ-BCN (-1.99 eV) < γ-BCN (-1.94 eV) < α-BCN (-1.87 eV) < θ-BCN (1.78 eV) < β-BCN (-1.70 eV). These stabilizing energies suggest that the most energetically favourable phase is η-BCN. It has also been found that the possibility of sustainability in the free-standing environment decreases from η-BCN to β-BCN phases. Furthermore, atomic bonds would describe the thermodynamic stability difference in the hybrid bpn-BCN phases. The C-C, C-N and B-N bonds have higher bond strength than N-N, C-B, and B-B in the crystalline geometry.[69] As a result, the phase with a large number of C-C and B-N bonds would be the most stable. According to **eq. (1)** and **eq. (2)**, the η-BCN phase has the higher magnitude of cohesive energy out of all the phases, and it is appreciable because it includes one C-C bond and four B-N bonds out of a total of nine bonds per unit cell. The number of B-N bonds in the remaining phases is constant, as reported in **Table 1**. Similarly, after η-BCN phases, δ-BCN has four C-N bonds, which shows that after the B-N bond, C-N bonds are more favourable than B-B and B-C bonds others.

*Table 1 Reports the formation and cohesive energies of bpn-BCN phases and their respective types and number of bonds per unit cell.*

| Phases | Energies(eV/atom) | | Types of Bonds and their counts per unit cell | | | | | |
|---|---|---|---|---|---|---|---|---|
| p-BCN | $E^{coh}_{p-BCN}$ | $E^{form}_{p-BCN}$ | B-B | C-C | N-N | C-N | B-N | B-C |
| α-BCN | -1.87 | 0.27 | 1 | 1 | 1 | 2 | 2 | 2 |
| β-BCN | -1.70 | 0.44 | 1 | 1 | 1 | 2 | 2 | 2 |
| γ-BCN | -1.94 | 0.34 | 1 | 1 | 1 | 2 | 2 | 2 |
| δ-BCN | -1.99 | 0.15 | 1 | 0 | 0 | 4 | 2 | 2 |
| η-BCN | -2.31 | -0.16 | 0 | 1 | 0 | 2 | 4 | 2 |
| θ-BCN | -1.78 | 0.18 | 0 | 0 | 1 | 2 | 2 | 4 |

In summary, the stability of hybrid bpn-BCN phases is determined by the total number of C-C, B-N, and C-N bonds. The η-BCN phase, characterized by stronger C-C and B-N chemical bonds, emerges as the most stable within the bpn-BCN family.

### 3.2.2. Phonon dispersion calculations

The phonon dispersion has been determined to assess the dynamical stability of bpn-BCN phases. **Fig. S3** depicts the phonon dispersion plots of the several bpn-BCN phases under investigation. The phonon branches cleanly miss imaginary frequencies, indicating the dynamical stability of the bpn-BCN phases. The highest optical phonon branches are obtained at frequencies 47.68, 44.23, 46.31, 46.61, 44.83, and 48.1 THz for α-BCN, β-BCN, γ-BCN, δ-BCN, η-BCN and θ-BCN, respectively. These small distinguish in the highest frequencies may be due to different bonding strengths between the atoms of these isoelectronic structures. In a nutshell, all phases of the hybrid bpn-BCN monolayer are equally probable to be dynamically stable.

### 3.2.3. Ab-initio molecular dynamics simulations

The thermal stability of bpn-BCN monolayers has been examined using AIMD trajectories at 300 K and 800 K at 10 *ps* with a time step of 1*fs*. **Fig. S4** demonstrates snapshot and free energy vs time AIMD simulations of bpn-BCN monolayers at 300 K temperature. The AIMD trajectories show that all phases could remain intact at 300 K with highly uniform temperature and energy profiles, demonstrating the thermal endurance of the bpn-BCN monolayers. Each upper panel of **Fig. S4** also depicts the original and final structure of the corresponding phases at 300K.

At 800K, a few of the atoms slightly buckled out of the plane but remained coordinated with the neighbours, showing the bond flexibility in **Fig. S5**. It is worth noting that trans-phases exhibit the least out-of-plane atom buckling, whereas cis-phases show a significant buckle height. This result will surely motivate researchers in the experimental production of 2D bpn-BCN monolayers.

### 3.2.4. Mechanical strength

The mechanical behaviour and stability of the different phases of bpn-BCN monolayers have been determined by computing its linear elastic constants using harmonic approximation. The elastic coefficients validated the Born criteria[70]: $C_{11}$, $C_{12}$, $C_{22}$ and $C_{44} > 0$ & $C_{11}C_{22} > C_{12}^2$ for phases of bpn-BCN monolayer. Additionally, the mechanical properties of the BCN monolayers are estimated, namely the in-plane Young's modulus $Y(\Phi)$ and Poisson ratio $\upsilon(\Phi)$. The elastic constants may be used to describe $Y(\Phi)$ and $\upsilon(\Phi)$ as functions of angle ($\Phi$) relative to the x-axis direction as follows:

$$Y(\Phi) = \frac{C_{11}C_{22} - C_{12}^2}{C_{11}s^4 + C_{22}c^4 + \left(\frac{C_{11}C_{22} - C_{12}^2}{C_{44}} - 2C_{12}\right)c^2s^2} \tag{3}$$

$$\upsilon(\Phi) = -\frac{\left(C_{11} + C_{22} - \frac{C_{11}C_{22} - C_{12}^2}{C_{44}}\right)c^2s^2 - C_{12}(s^4 + c^4)}{C_{11}s^4 + C_{22}c^4 + \left(\frac{C_{11}C_{22} - C_{12}^2}{C_{44}} - 2C_{12}\right)c^2s^2} \tag{4}$$

where $c = cos\Phi$ and $s = sin\Phi$ based on calculated elastic constants. The angle-dependent $Y(\Phi)$ and $\upsilon(\Phi)$ are shown for different phases has been shown in **Fig. S6**.

It is found that the angular dependence of Y varies in the range of 143.93 to 253.62 N/m for different phases of bpn-BCN monolayers, showing that Y is highly anisotropic in nature. The maximum values of Y are comparable to our previous work, a carbon biphenylene network BPN (256.21 N/m) and an identical boron-nitride network I-BPN (193.67 N/m).[55] Moreover, Poisson's ratio has an average value larger than 0.34, indicating that bpn-BCN phases have a lesser brittle potential.[71] If a comparison has been made within the different phases, γ-BCN shows the highest Y values, whereas α-BCN and δ-BCN show identical responses, as shown in **Fig. S6**. The extremum of Y and υ has been reported in **Table S2**. The η-BCN phase has the greatest young-modulus anisotropy with a difference of 72.13 N/m in the transverse direction, whereas α-BCN demonstrates the least difference, 51 N/m. In summary, the high Young's moduli of all phases signify their capacity to resist elastic deformation, while the positive Poisson ratio indicates their ability to expand or contract in response to compressive or tensile strain in opposing directions within the phases.

### 3.3. Electronic properties

Further, the electronic properties of bpn-BCN phases have been investigated. **Fig. 2** displays the electronic band structures along the high-symmetry k-path and the projected density of states, which show the atomic contribution, especially around the Fermi level, in different phases of bpn-BCN. It is observed that four (β- BCN, γ- BCN, δ-BCN and θ-BCN) out of six phases are all metallic because the bands cross the Fermi level. α-BCN is semi-metallic as because of valence bands maximum (VBM) and conduction bands minimum (CBM) just touch the Fermi level, and η- BCN phases is a semiconductor with a small direct band gap of 0.19 eV. In addition, cis-phases of bpn-BCN structures (three in the top panel of **Fig. 2**) exhibit single band crossing, whereas, in trans-phases (bottom panel of **Fig. 2**), it can be seen that more than one bands cross the Fermi-level. Interestingly, in the η-BCN phase, the bands approach the Fermi level from both ends at high symmetry C point, the valence band with sharp V shape dispersion while the conduction band approach the Fermi energy at C point and goes to flat over the symmetry path. This η-BCN phases has narrow bandgap of 0.19 eV. The projected density of states shows the atomic contribution at different energy levels. The α-phase shows that the major contribution at Fermi energy comes from C atoms than N atoms and a small contribution from B atoms. These hybrid bands slightly touch the Fermi energy. The β-phase possesses a dominant contribution from N rather than C and B atoms.

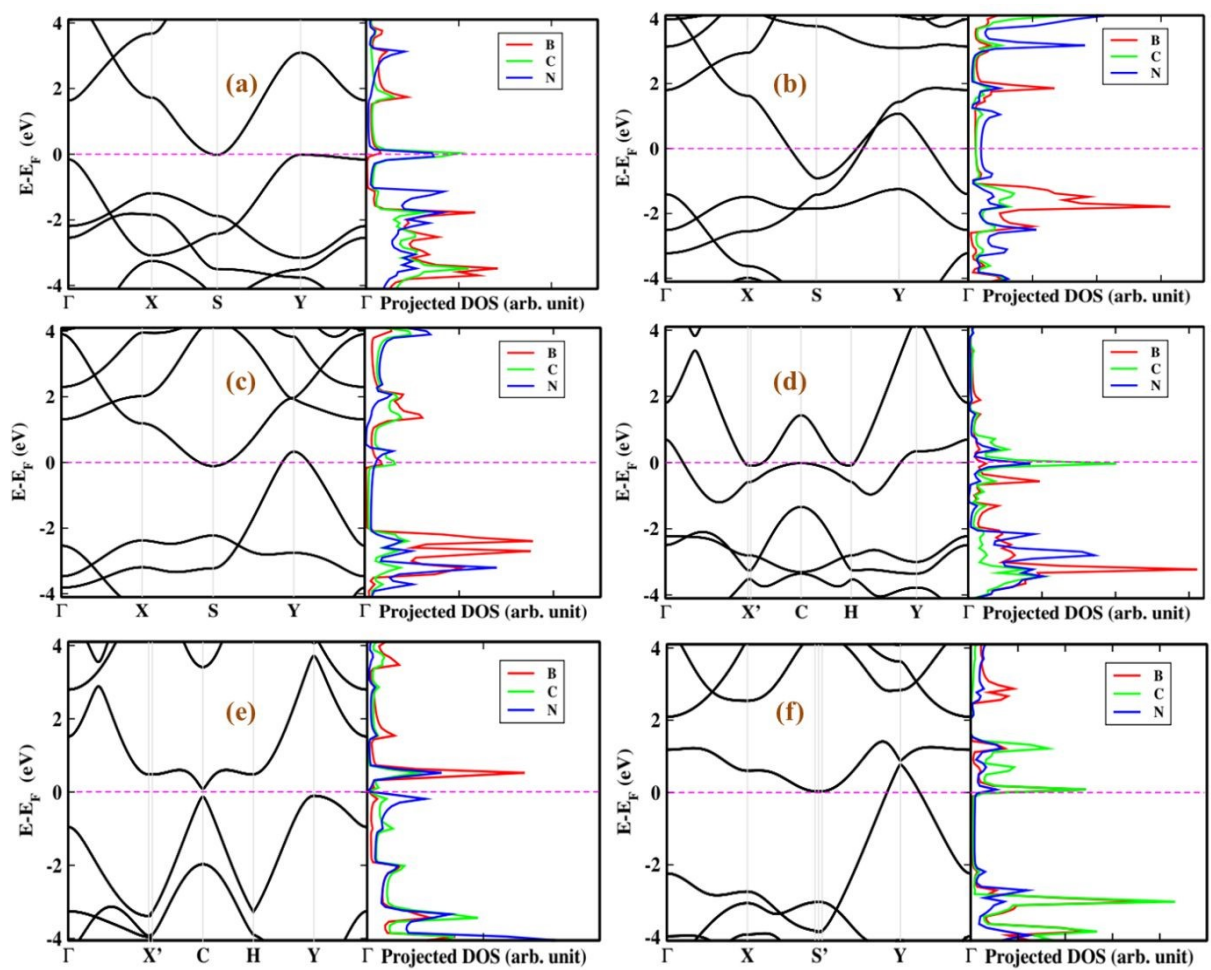

*Fig. 2* Shows the electronic bands structures (left panel) and projected density of states (right panel) for different phases of bpn-BCN monolayers: α- BCN, β- BCN, γ- BCN, δ-BCN, η- BCN and θ-BCN are marked in the figure as a, b, c, d, e and f respectively.

Similarly, in the γ-phases, there is a mixture of p-orbitals of B, C, and N at the Fermi energy. In the δ- and θ-phases, p-orbitals of C atoms dominate, exhibiting sharp (green) peaks in the PDOS plots. However, in the η-phase, B and N show a major contribution in the conduction bands, with dominance of N orbitals in the valence bands in the vicinity of the Fermi line. Also, it should be noted that the main contribution of the p-orbital of each atom (especially $P_z$) dominates near the Fermi level. To verify the robustness of the metallic nature of bpn-BCN phases, alongside one semiconductor, the Heyd-Scuseria-Ernzerhof screened hybrid functional has also been employed. In **Fig. S7**, the HSE electronic band structures are displayed. It has been seen that the dispersion of electronic bands varies across the energy range, while the metallic character of the bpn-BCN phases remains unchanged. Notably, the band gap for η-BCN phases is determined to be 0.68 eV, highlighting the accuracy of the HSE technique in capturing electronic properties. These findings underscore the consistency of the metallic properties of bpn-BCN phases and overcome the underestimation inherent in the GGA-PBE functional, providing valuable insights into their electronic behavior.

### 3.4. Possibility of experimental synthesis of bpn-BCN

There are primarily two methods for creating 2D materials. The top-down method involves extracting a 2D monolayer from its bulk phase. For example, Nobel laureates A.K. Geim and K. Novoselov extracted graphene, a single-atom-thick crystallite (graphene), from bulk graphite using the Scotch Tape Method. This innovative top-bottom approach involved transferring graphene onto a silicon wafer coated with thin silicon dioxide.[72]

Another option is the bottom-up approach, in which small molecules or atoms are put together in a pattern under specific circumstances using sophisticated technology, a nonreactive substrate, and other essential factors. The molecular beam epitaxy (MBE), chemical vapour deposition (CVD), and sol-gel method, among many others, are examples of this approach.[73] Graphene-like network of boron nitrogen synthesis by CVD is one of the bottom-up approaches for synthesizing 2D materials. The synthesis of h-BN films on a copper substrate with large (20 μm) single crystals using borazine as a precursor at atmospheric pressure.[74] Ci, L. et *al.* were the first to report synthesizing the BCN film on Cu substrate using the thermal catalytic CVD technique. Methane for C and ammonia borane ($NH_3$-$BH_3$) for BN were precursors. It is also reported that modifying the experimental conditions may change the atomic ratios of B, C, and N. However, the equal stoichiometric ratio of B and N had been maintained in the synthesized thin film.[75] Recently, g-BCN thin film has been synthesized using bis-BN cyclohexane[76], where $B_2N_2C_2H_{12}$ was used as a precursor molecule. Moreover, the dehydrogenation precursor molecules and covalent bond formation will be thermally driven on an epitaxial monolayer on Ir(111).[40] Let's talk about the experimental procedure used to synthesize g-BCN by using 1,2-BN cyclohexane as a precursor. The molecules 1,2-BN cyclohexane were sublimated at room temperature in an ultra-high vacuum pressure range. This sublimated molecular powder was put in a quartz crucible for further procedure to generate molecular vapour at room temperature. After cleaning the Ir(111) substrate with Ar+ ion sputtering, various samples were prepared by molecule deposition using e-beam heating.

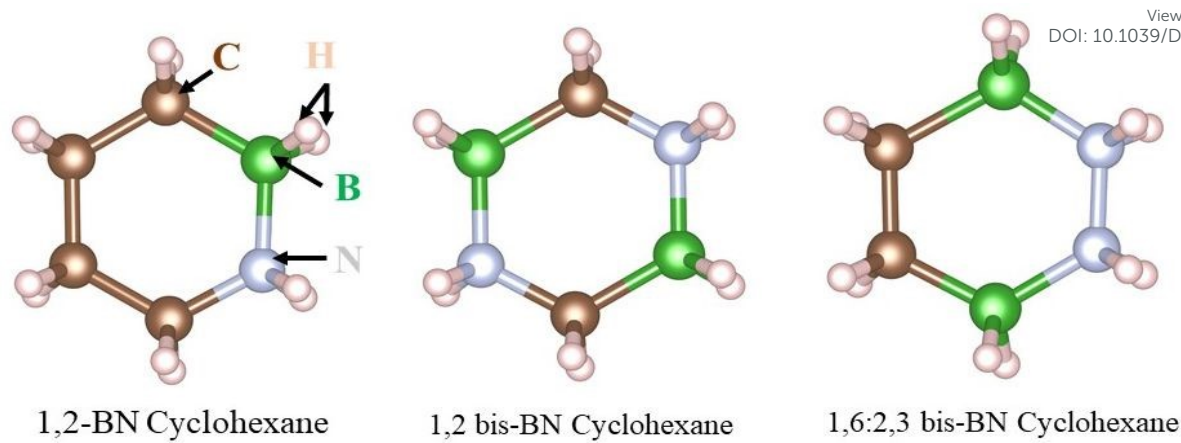

*Fig. 3* *Shows the molecules that can be precursors to synthesize the bpn-BCN monolayer.*

In a similar manner, it may be possible to synthesize the various phases of the hybrid bpn-BCN monolayer. There are three main precursors that will help to synthesize the bpn-BCN experimentally, as shown in **Fig. 3(a)** 1,2-BN cyclohexane,[77] **3(b)** bis-BN Cyclohexane,[78] and **3(c)** 1,6;2,3-bis-BN cyclohexane.[76] All these precursors have been experimentally reported, including 1,6;2,3-bis-BN cyclohexane, which was recently synthesized by Dai et al. The various phases of the bpn-BCN monolayer can be created with freshly synthesized 1,6;2,3-bis-BN cyclohexane. The dehydrogenated precursor molecule may operate as a free radical and interact with other molecules. As a result, growing a molecular cluster or chain results in the formation of a 2D network. All phases of bpn-BCN might be feasible depending on the interaction location of either the separate precursor 1,2-BN cyclohexane or 1,6;2,3-bis-BN cyclohexane or both combined in the thin film deposition.

### 3.5. η-BCN as an anode material for Alkali metal ions batteries.

The superiority of choosing 2D materials for anode applications lies in their high surface-to-volume ratio, which enhances the interaction with adsorbed metal atoms. The reduction of one dimension results in several dangling bonds, increasing surface energy. Additionally, the ionic conductivity of alkali ions over 2D materials is superior to that of bulk systems. These advantages are beneficial for studying the microstructure in electrochemical applications. Structural stability is crucial for the performance of electrode materials in electrochemical applications, as the lifespan of a battery depends on the durability of the electrode material through numerous charging and discharging cycles. To verify this stability, we conducted various calculations, including formation energy for thermodynamic stability, phonon dispersion for dynamic stability, Young's modulus for mechanical strength, and AIMD simulations at finite temperatures.

Here, we preferred the η-phase of the hybrid bpn-BCN monolayer for electrochemical properties for alkali metal ions batteries because it is the most stable phase of hybrid bpn-BCN. In the subsequent sections, we exclusively refer to the η-phase of bpn-BCN as simply "bpn-BCN" for our electrochemical investigation. As discussed earlier, the geometry of the η-BCN monolayer is oblique, with a = 3.86 Å, b = 4.55 Å, and α= 92.5°. As illustrated in **Fig. 1(e)**, the bpn-BCN has a single-layer geometry with 2B, 2C, and 2N atoms per unit cell arranged in hexagons, octagons, and square rings. Its remarkable stability compared to other phases is because it contains the highest number of C-C and B-N bonds. The primary reason for the strength of these C-C and B-N bonds is that they satisfy the octet or duplet rule, ensuring that each atom has a complete set of electrons in its outermost shell. Furthermore, there are probably C-C, B-N, C-N, and B-C bonds with three coordination numbers for each atom, indicating $sp^2$ hybridization bpn-BCN monolayer. Although the charge distribution is not uniform, the π-cloud was not discovered over the bonds because B-N bonds are partially ionic in nature. The electron localization function plot in **Fig. S2(e)** shows the same result. It has been observed that more charge accumulates in the region of the N atom, and a deficiency in charge at the B atom confirms the partial ionic character of the B-N bond. There is no observable charge polarization in the C-C and B-C bonds, indicating their covalent nature. As per the anode material required, the dynamics of atoms play an important role, so studying the phonon dynamics of the bpn-BCN monolayer is necessary. **Fig. S3(e)** shows a phonon dispersion plot that confirms the dynamical stability of bpn-BCN. Additionally, AIMD simulations were performed at high temperatures to investigate the thermal stability of the bpn-BCN monolayer. **Fig. S4(e)** displays the overall energy time evolution during the simulation and a snapshot of the structure calculated at 300 K with a time scale of 1 *fs* for a time period of 10 *ps*. There have been no structural reconstructions or broken bonds observed. The planarity of the bpn-BCN monolayer is preserved despite a small displacement of atoms from their mean positions.

The conductivity of anode material is particularly important because electrons and ions must pass between the electrode and electrolytic components to complete the charging and discharging cycles of batteries. Therefore, the material with a metallic character is always preferred as an anode material. Additionally, the cyclability and rate performance of batteries significantly relate to the electronic properties of electrode materials. Hence, the electronic properties have been investigated by calculating the electronic band structure and projected density of states of the bpn-BCN monolayer using PBE-GGA exchange-correlation functional. The band structure in **Fig. 2(e)** reveals that the bpn-BCN monolayer is a semiconductor with a small direct bandgap of 0.19 eV. According to the projected density of states (PDOS) in **Fig. 2(e)**, the valence band predominantly comprises N atoms. In contrast, the conduction state is a mixture of the p-orbital of B and N atoms. Furthermore, the $p_z$ orbital contributes significantly to the contribution of N and B atoms in the bands nearing Fermi energy. In addition, the conduction band is flat somewhere around X' and Y while passing through the C-high symmetry point. The degeneration in conduction bands helps to adsorb the alkali metal atoms on the conduction site of the bpn-BCN monolayer.

The choice of bpn-BCN for electrochemical studies is particularly advantageous. First, its structural stability, high Young's modulus, and structural sustainability up to 800K indicate it can withstand many cycles. Secondly, the bpn-BCN monolayer can adsorb a significant number of alkali metal atoms due to its various shape motifs, which result in high surface energy. The BCN combination includes partially ionic (B-N, B-C) and covalent bonds (C-C), creating various possible adsorption sites. Lastly, the presence of different shape motifs further facilitates the adsorption of alkali metal atoms on the bpn-BCN monolayer.

### 3.5.1. Adsorption of alkali metal atoms

The adsorption behaviour of alkali metal atoms (M=Li, Na and K) on the bpn-BCN has been investigated by finding the potential adsorption sites for metal ions on the monolayer. For this, six different sites have been considered: H-site (centre of hexagon site), S-site (centre of square site), O-site (centre of octagon), BN-site (centre of B-N bond), CC-site (centre of C-C bond) and BC-site (centre of B-C bond) as shown in **Fig. S8.** The geometry optimization revealed that only three of these sites, H-site, S-site, and O-site, are stable, as shown in **Fig. 4.** The metal atom on the other sites shifted to any stable sites during the optimization. The adsorption energy ($E_{ads}$) of the stable sites has been calculated by taking the difference between the total energy of the M-adsorbed bpn-BCN system and the sum of the total energy of pristine bpn-BCN monolayer and the energy of isolated alkali metal (M). Mathematically, it is defined as

$$E_{ads} = E_{M@bpn-BCN} - E_{bpn-BCN} - E_M \;; (M = Li, Na, K) \qquad (5)$$

where $E_{M@BCN}$ is the total energy of metal adsorbed bpn-BCN monolayer, $E_{BCN}$ is the energy of pristine bpn-BCN monolayer and $E_M$ is the energy of isolated single metal atom. The adsorption energy has been calculated using a supercell of 3×3×1 to neglect the interaction between the alkali metal ion. Our global search for stable adsorption confirmed that the H-site is the most stable, followed by O- and S-site for all alkali metals.

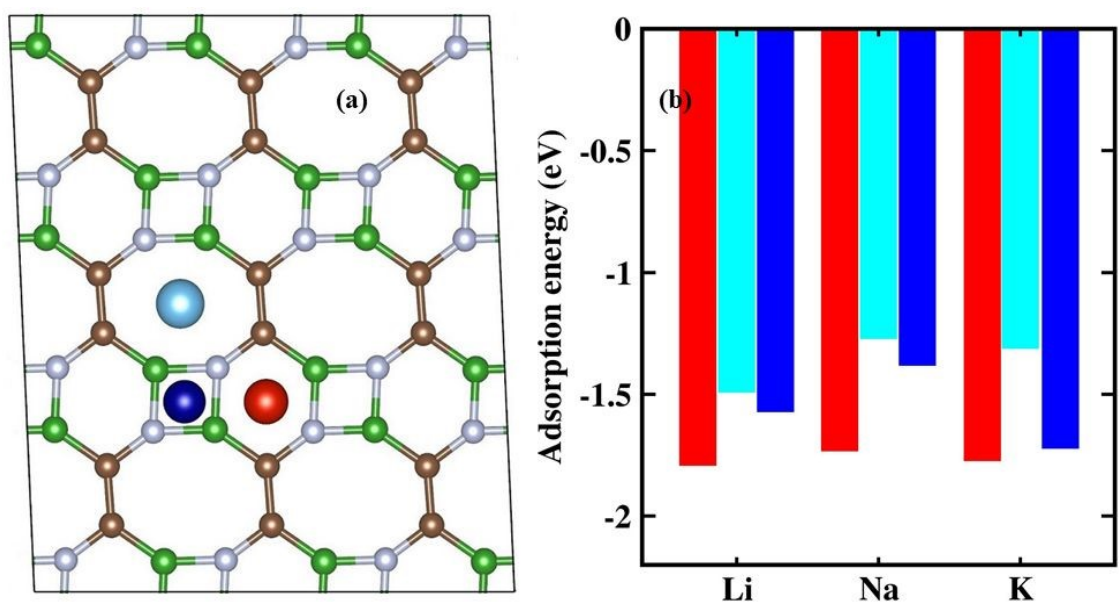

*Fig. 4 The adsorption energies of single alkali metal atoms (M = Li, Na, K) on the hybrid bpn-BCN monolayer at the three stable adsorption sites H-site, O-site, and S-site as shown in red, cyan and blue colour respectively.*

Furthermore, Li shows better adsorption in all respective sites compared to Na and K. Its small atomic size makes it easier to be adsorbed rather than charge transfer. In contrast, charge transfer for K is higher than Li and Na because of the screening effect caused by core electrons with large radii. It has been confirmed that the H-, O-, and S-sites favourably adsorbed the metal atoms due to their negative adsorption energies as reported **in Fig. 4** (and in **Table 2)**. Still, the H-site is the most favourable as the magnitude of adsorption energy is large value.

*Table 2 The adsorption energy, the distance between the metal atom and bpn-BCN monolayer (Height), and the charge transfer at different sites for alkali metal ions, respectively.*

| Alkali Metal | $E_{ads}$ (eV) | | | Height (Å) | | | Charge transfer (e) | | |
|---|---|---|---|---|---|---|---|---|---|
| | H-site | O-site | S-site | H-site | O-site | S-site | H-site | O-site | S-site |
| Li | -1.79 | -1.49 | -1.57 | 1.66 | 1.68 | 1.82 | 0.47 | 0.41 | 0.45 |
| Na | -1.73 | -1.27 | -1.38 | 2.10 | 1.91 | 2.21 | 0.56 | 0.51 | 0.53 |
| K | -1.77 | -1.31 | -1.72 | 2.51 | 2.37 | 2.56 | 0.61 | 0.58 | 0.57 |

The COHP analysis of the adsorption process reveals crucial insights into the interactions between alkali metals and the bpn-BCN monolayer. Projected COHP (pCOHP) has been calculated to examine the interactions of foreign alkali metals with the local atoms of the bpn-BCN monolayer. In this analysis, a positive pCOHP value indicates anti-bonding interactions, while a negative value indicates bonding interactions, as shown in **Fig. 5**. The pCOHP plots for Li are shown in **Fig. 5**, while the plots for Na and K are presented in **Figs. S9 and S10**, respectively. For the H-site, the pCOHP value for B-Li shows a peak at the Fermi energy, indicating strong bonding orbitals that stabilize the adsorption process, C-Li and N-Li interactions exhibit only small spikes at the Fermi energy for the H-site, and there are no significant spikes for the O-site and S-site. For the S-site and O-site, the pCOHP of N-Li shows bonding, which seems to favour adsorption. Similarly, pCOHP, for Na and K shows weak van der Waals interactions with bpn-BCN. To further clarify these interactions, we integrate the pCOHP from negative infinity to the Fermi level (ICOHP) which represents interaction intensity, more negative ICOHP values indicating stronger bonding and more positive values indicating stronger antibonding.[79, 80] **Table S3** reports the ICOHP values, showing that more negative ICOHP values for N and B with alkali metals suggest stronger bonding for S- and O-site and C with alkali metal for H-site. For a given pair of atoms, a more negative ICOHP value indicates stronger bonding interactions. The integrated pCOHP values are reported in Table S3. These values suggest that the B and N atoms interact more with the alkali metals, facilitating adsorption. As shown in Table S3, the ICOHP values follow the C-M > B-M > N-M for the O-site and S-site for all alkali metals (M = Li, Na, K), indicating that N-M interactions are more favourable (more -ve) in these sites. However, for the H-site, the trend is different: B-M > N-M > C-M for Li, Na, and K, indicating that C-M interactions are more favourable for the adsorption process in this site.

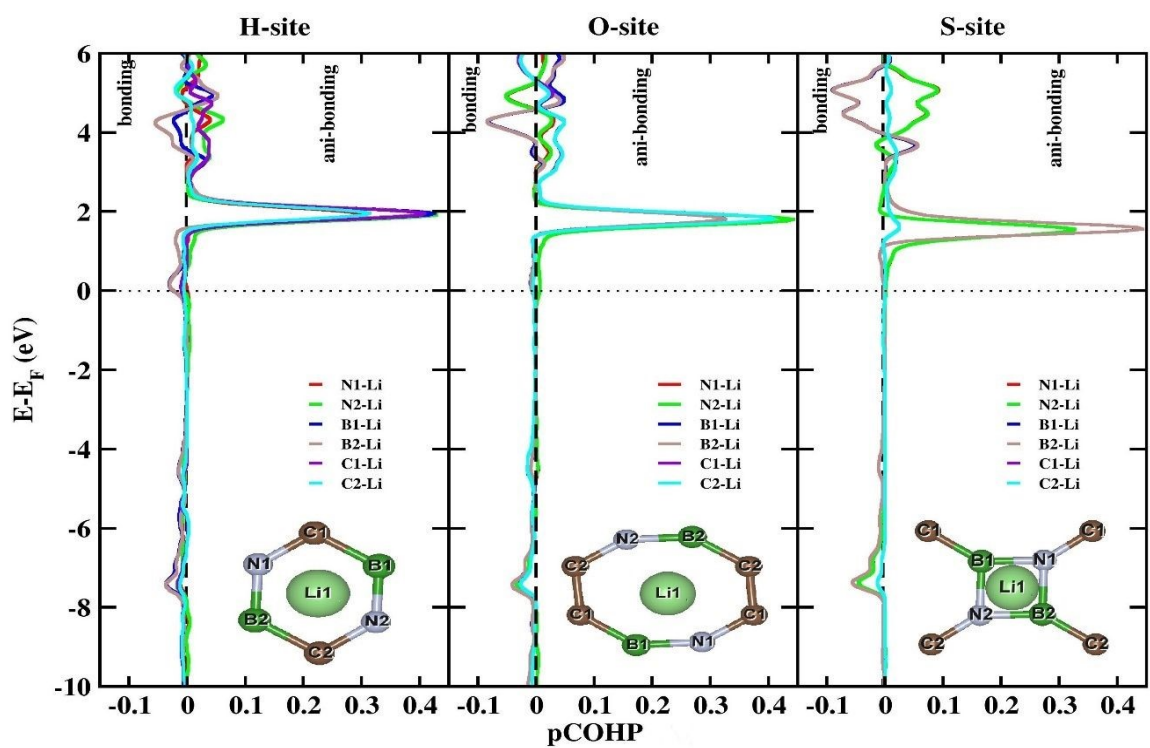

***Fig. 5*** *COHP analyses of the chemical bonding between the adsorbed Li atom and constituent atoms of the bpn-BCN monolayer for (a) H-site, (b) O-site, and (c) S-site. The bonding and antibonding states are indicated by negative and positive values of pCOHP respectively.*

In order to check the conductivity of the metal-adsorbed bpn-BCN monolayer, the electronic band spectra have been examined for all favourable sites. The charge transfers from alkali metal to bpn-BCN monolayer, making it metallic. The electronic band structures of Li-adsorbed at different sites of bpn-BCN monolayer have been illustrated in **Fig. 6.** With the increase in alkali metal concentration, more electrons will enter the conduction bands, and hence there is an increase in conductivity of metal-adsorbed bpn-BCN monolayer.

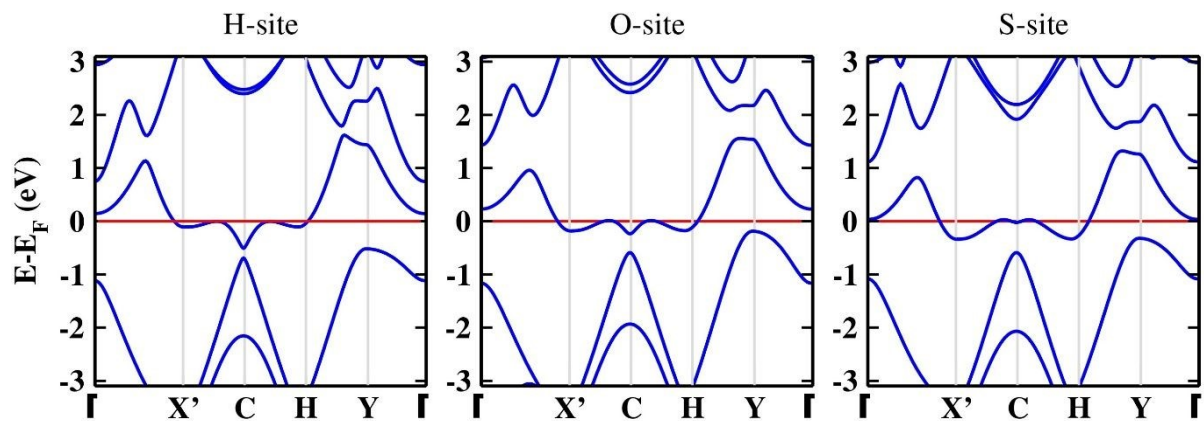

*Fig. 6 The electronic band spectra of Li metal adsorbed bpn-BCN monolayer at H- O- and S-sites.*

Similarly, the Na- and K-adsorption also maintain the metallic nature of the bpn-BCN monolayer, as shown in **Fig. S11** and **Fig. S12**. Therefore, the lithiation process of bpn-BCN monolayer demonstrates improved electrical conductivity, suggesting that there is no need for additional conductive additives when it functions as electrodes.

### 3.5.2. Electron density difference

The charge redistribution has been investigated using electron density difference analysis to understand more about the metal-adsorbed bpn-BCN monolayer. The following formula has calculated the electron density difference between pristine and alkali-metal adsorbed bpn-BCN monolayer.

$$\rho_{net} = \rho_{(M@bpn\text{-}BCN)} - (\rho_{bpn\text{-}BCN} + \rho_M) \tag{6}$$

Where $\rho_{(M@bpn\text{-}BCN)}$ and $\rho_{bpn\text{-}BCN}$ are the electron densities of the alkali-metal adsorbed bpn-BCN monolayer and pristine bpn-BCN monolayer, respectively, and $\rho_M$ is the electron density of alkali metal atom. **Fig. 7** shows electron density difference plots of Li adsorbed on a bpn-BCN monolayer at three distinct stable sites, demonstrating the change in electronic distribution produced by charge transfer from alkali metal to bpn-BCN. The average planar electron density difference displayed the distribution of charge density difference along the vertical direction, with the bpn-BCN monolayer placed at the centre of the box (z = 10 Å), as shown in the **Fig. 7**. This analysis visualizes the redistribution of electron density difference upon adsorption of alkali metals, highlighting regions of charge accumulation (yellow) and depletion(cyan) above and below the monolayer. The alkali metal and bpn-BCN monolayer interact strongly through coulomb forces, leading to high adsorption energy.

Nanoscale  Page 16 of 26

16
View Article Online
DOI: 10.1039/D4NR01386G
Published on 07 June 2024. Downloaded by Indian Institute of Technology Patna on 6/16/2024 12:57:47 PM.

Nanoscale Accepted Manuscript
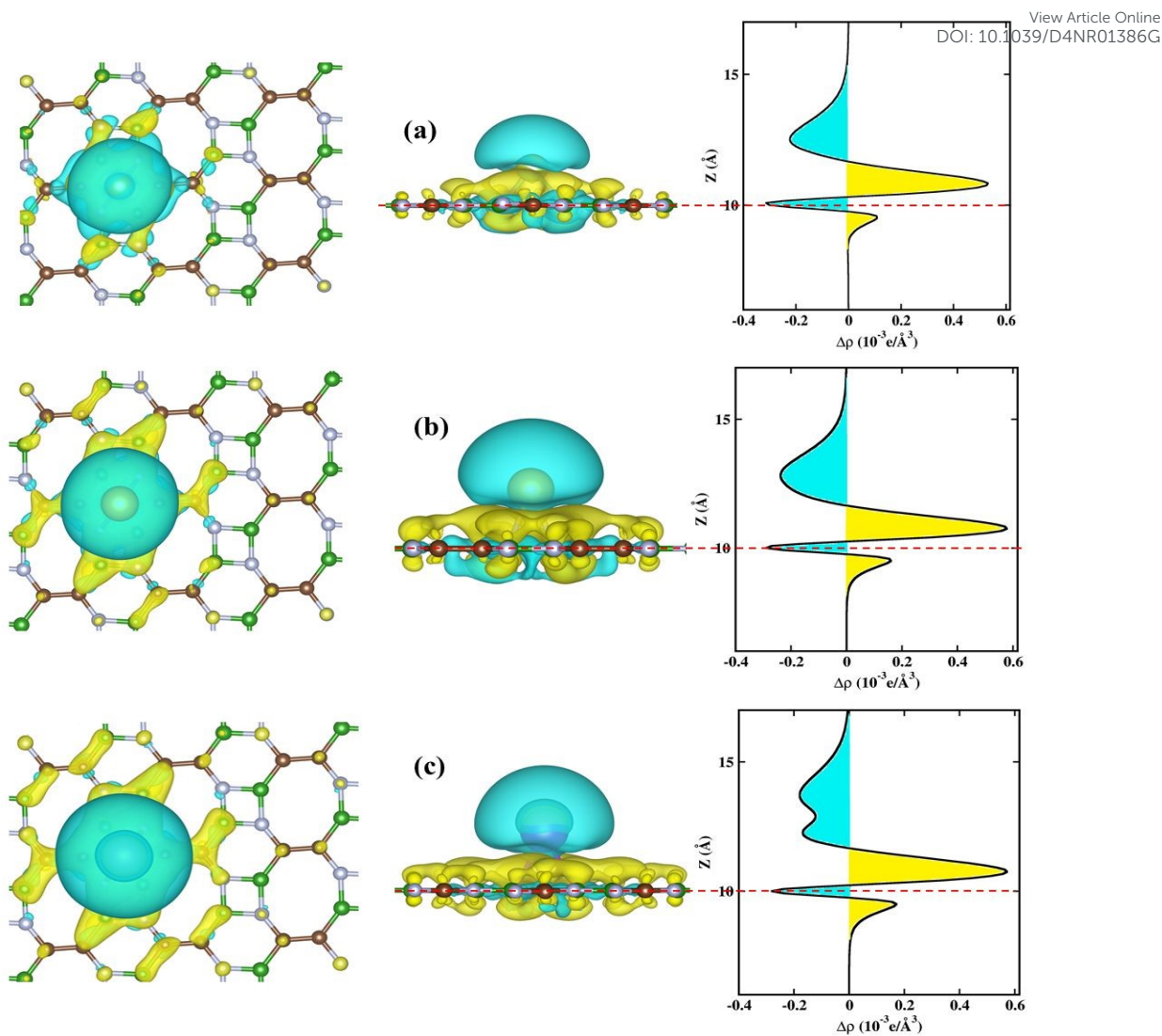

*Fig. 7* *The top and side view along with average planar electron density difference for (a) Li- (b) Na- and K-metal adsorbed at (a) H-site on the bpn-BCN monolayer with iso-surface value of 0.001 eÅ$^{-3}$.*

This interaction caused a significant electron depletion around the alkali metal and electron accumulation at the bpn-BCN surface, as seen in **Fig. 7**. The electron density difference plots for S- and O- adsorption sites for alkali metals are displayed in **Fig. S13** and **Fig. S14**. To quantify the charge transfer, we have calculated Bader atomic charge on each atom. According to the Bader charge analysis, charge transfer occurs more for K than Na and Li, and its values depend on the adsorption site. At a specific site, such as the H-site, the value of charge transfer varies among alkali metal atoms, with K exhibiting the highest transfer and Li the lowest. It can be explained by the fact that core electrons screened the valence electrons due to increased atomic radius. Similarly, at the S-site, charge transfer occurs from metal atoms to bpn-BCN due to the unique electron-deficient (B) and excess (N) interactions experienced by alkali metals at the diagonal of the square motif. On the other hands, H- and O-site metal atoms face different interactions due to B-N, C-C bonds, and their spatial distance from the centre of the motifs, resulting in a slight change in charge transfer. It has been noticed that the charge transfer values are directly related to the adsorption energies of alkali metal atoms on the bpn-BCN monolayer. The charge transfer from Li to bpn-BCN monolayer at H-site (0.47e$^-$), S-site (0.45e$^-$) and O-site (0.41e$^-$) show the H-site is more favourable in the adsorption of Li on the bpn-BCN monolayer. Na and K metal atoms have noticed a similar trend for different adsorption sites. The value of charge transfer, adsorption height and adsorption energy for Li/Na/K at different stable sites has been reported in **Table 2.**

### 3.5.3. Diffusion of alkali metal atoms on the bpn-BCN monolayer

To estimate the ionic conductivity of the bpn-BCN monolayer, we rely on the diffusion barrier energy to approximate the diffusion coefficient. For this, it has been essential to identify the minimal energy path along the high symmetry direction. Three different high-symmetry diffusion paths have been chosen to investigate the diffusion barrier energy. As shown in **Fig. 8(a)**, all diffusion paths start from a highly stable H-site and end at another H-site by passing to the S-site and O-site in path 1 and path 2, respectively, whereas in path 3, ion diffused between two H-site by passing through the C-C bond. Path 2 has the lowest diffusion barrier for all alkali metal ions, and path 3 has the highest barrier height, as shown in **Fig. 8(b-d)**. Li encounters the most difficulty along Path 2 with a barrier height of 0.65 eV compared to Na 0.26 eV and K 0.23 eV. The higher barrier of Li in contrast to Na and K ions is due to its high adsorption and experiencing the potential depth at the H-site than the entire BCN surface. This diffusion barrier for Li is slightly higher than graphene (0.32)[81] but has a small value as compared to other 2D materials like boron sheets(0.79).[26] The carbon biphenylene network reported a diffusion barrier of 0.23 eV[82] for Li but has small adsorption energies because of all identical covalent C-C bonds. Further, the diffusion barriers for Na and K ions along all three paths are easily feasible and comparable with the other BCN hybrid monolayers.[57, 83]

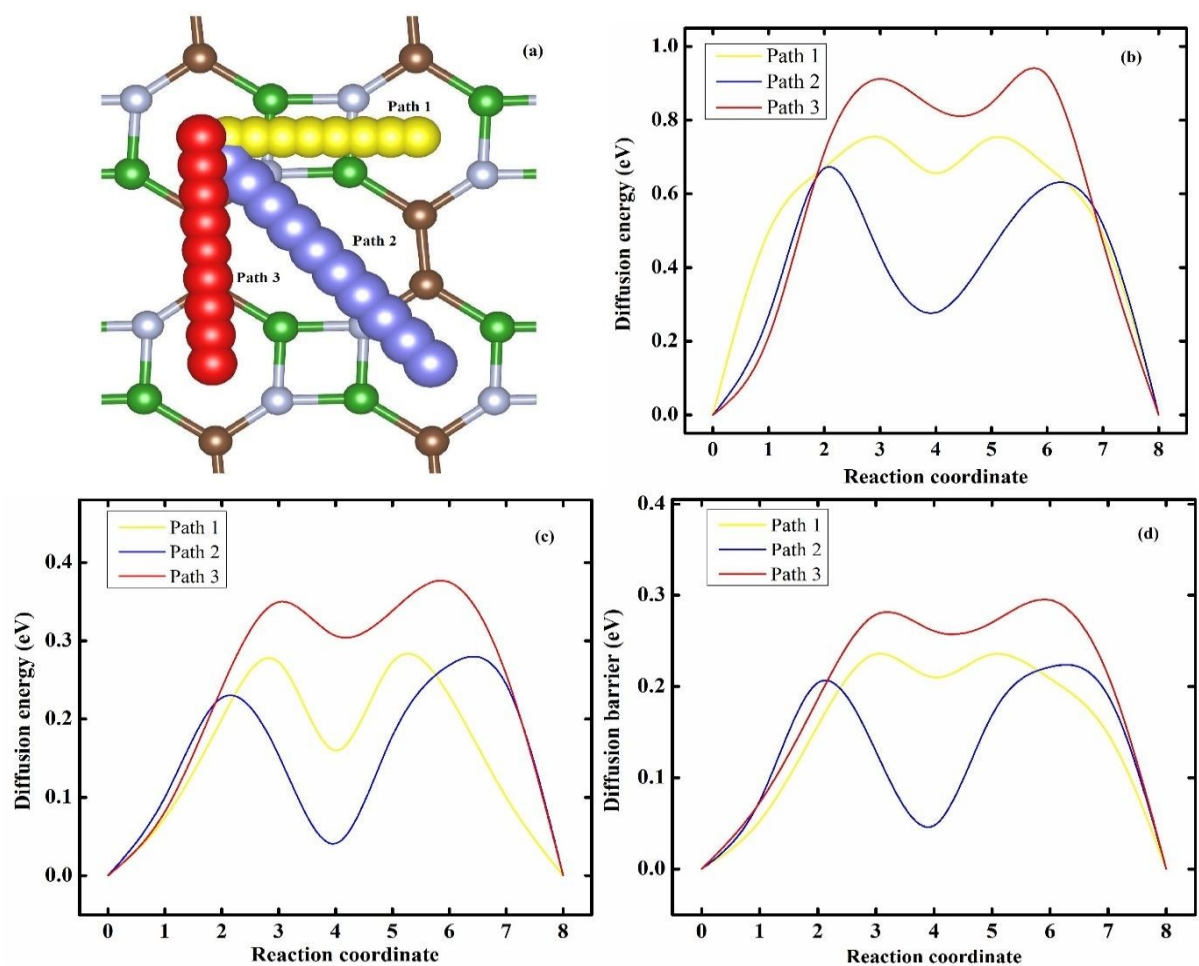

***Fig. 8***. *Displays (a) the possible diffusion paths and the corresponding diffusion energies for (b) the Li, (c) the Na, and (d) the K atom.*

### 3.5.4. Open circuit voltage and theoretical storage capacity

The average open circuit voltage (OCV) and storage capacity (C) of rechargeable metal-ion batteries are critical parameters that must be determined for the charge-discharge mechanism. The OCV theoretically can be derived by calculating the average voltage across different alkali metal composition domain. The following reaction has been used to illustrate the charge-discharge mechanism of a bpn-BCN monolayer based metal-ion battery:

$$E_{M_{x_1}BCN} \rightarrow E_{M_{x_2}BCN} + (x_2 - x_1)E_M \qquad (7)$$

The OCV has been estimated by calculating the change in Gibbs free energy ($\Delta G$) for **eq. (7)** half-cell reaction[84].

$$\Delta G = \Delta E + P\Delta V - T\Delta S \qquad (8)$$

The second term in **eq. (8)**, P$\Delta$V, is negligible because of very small change in volume during the adsorption of alkali metal on bpn-BCN monolayer. Additionally, the entropy term (T$\Delta$S) in **eq. (8)** is roughly 25 meV at ambient temperature, which is very small compared to the average adsorption energy (1-2 eV) of alkali metals.[85, 86] Therefore, the change in Gibbs free energy is approximately equal to the change in internal energy ($\Delta$E), representing adsorption energy.

Hence, the average OCV of $M_xBCN$ in the range of $x_1 < x < x_2$ is calculated from the average adsorption energy,[87, 88]

$$OCV = -\frac{\Delta G}{(x_2 - x_1)ne} \approx -\frac{\Delta E}{(x_2 - x_1)ne} = -\frac{E_{ave}}{(x_2 - x_1)ne} \quad (9a)$$

$$E_{ave} = \frac{E_{M_{x_2}BCN} - E_{M_{x_1}BCN} - (x_2 - x_1)E_M}{(x_2 - x_1)} \quad (9b)$$

Where $E_{M_{x_2}BCN}$ and $E_{M_{x_1}BCN}$ are the energies of M-adsorbed bpn-BCN at two adjacent concentrations $x_2$ and $x_1$ respectively and $E_M$ is the energy of metal atom $M$. The symbol $'n'$ (where n=1 for alkali atoms) denotes the number of electrons that are completely ionized from alkali metals. The symbol $'x'$ represents the concentration of M atoms, and the maximum value of $x$ can be obtained by gradually increasing the metal atoms on the bpn-BCN monolayer.

The following equation has been used to compute the relevant adsorption capacities:[89]

$$C = \frac{xnF}{W_{M_xBCN}} \quad (10)$$

where $C, x, n, F$ and $W_{M_xBCN}$ are storage capacity, concentration of Metal atom per unit cell of bpn-BCN monolayer, number of valence state of metal atom, Faraday's constant and molecular mass of metal adsorbed bpn-BCN monolayer, respectively.

To determine the maximum storage capacity, metal atoms have been incrementally added to a monolayer of bpn-BCN, one at a time. For this, a 2×2×1 supercell ($B_8C_8N_8$) has been used to describe slowly adding metal atoms to the bpn-BCN surface. Adding a single atom of an alkali metal (denoted as M) to the H-site on either side of $B_8C_8N_8$ surface results in a stoichiometric ratio, represented as $x = 0.125$, within the $M_xBCN$ unit formula, reflecting the lower limit of alkali metal chemical content. Further studies show that M atoms in the H-site have an affinity to attach on both sides of a BCN monolayer. Once all the H-sites on two sides have been filled, M atoms occupy the O-sites in the second layer of M atoms. $M_xBCN$ system with higher $x$ values ($x$ = 0.125, 0.25, 0.5, 1, and 2) has been studied in a wide range of probable configurations with both surfaces exposed to metal atoms, as shown in **Fig. 9.** The same process has been used to create BCN systems with intercalated all alkali metal (Li, Na and K) atoms.

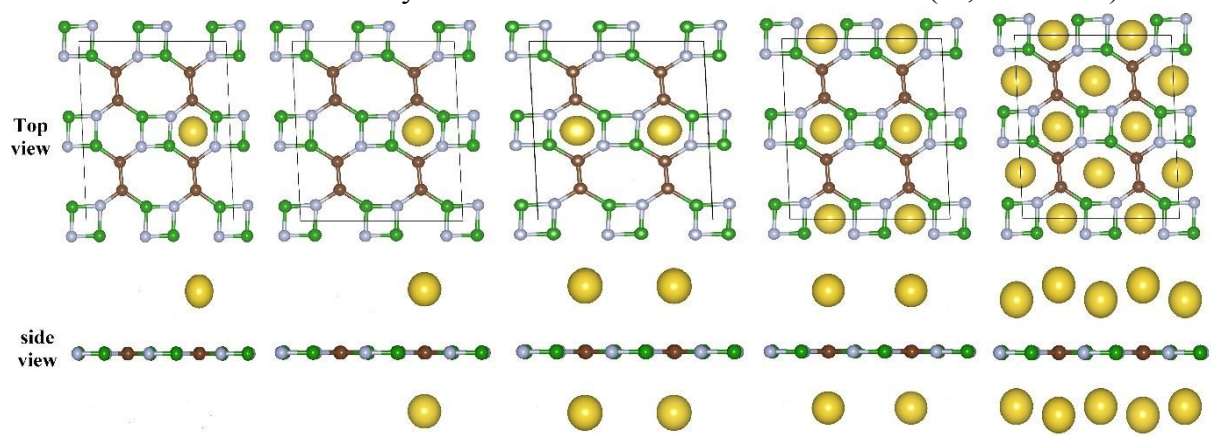

*Fig. 9* Displays the top (first panel) and side (second panel) view of optimized geometry by increasing the concentration of M atoms per cell ($B_8C_8N_8$) of bpn-BCN.

.

Further, to demonstrate the most stable adsorption configuration of metal atoms (M = Li, Na, K) at different adsorption concentrations on the bpn-BCN monolayer, we calculated the average adsorption energy ($E_{ave}$) of the $M_xBCN$ (M = Li, Na, and K) system for the pristine bpn-BCN monolayer and Li/Na/K metals as reference states using the **eq. 9(b),** which is further utilized for OCV calculations. **Fig. 10(a)** shows the average adsorption energies as the concentration of metal atoms increases. As the number of adatom layers increases, $E_{ave}$ should decrease to avoid creating metal atom clusters.[26, 90] A similar pattern has been observed in our alkali metal atoms calculations. Further, the decrease in $E_{ave}$ with concentrations of metal atoms is due the electrostatic repulsion among the adsorbed metal atoms. The adsorption energy decreases for all metal atoms up to four layers. Consequently, the maximum value of *x* for Li, Na, and K is four.

Finally, **eq. (9a)** and **(9b)** has been used to calculate the average OCV of alkali-ion batteries by varying *x* values. **Fig. 10(b)** depicts the estimated OCVs in $M_xB_8C_8N_8$ as a function of alkali atoms concentration. The storage capacity on the anode of all alkali-ion batteries grows as *x* increases, but the average OCV on the anode decreases. The average OCV for Li in commercial anode materials like graphite is 0.11 V,[91] and for $TiO_2$ is 1.50-1.80 V.[92] Similar to Na and K-ions, the insertion voltage V will be in the range of 0-1V. The average voltage for Na-ions in the graphite anode materials is 0.3 V,[93] and for K-ion in the graphite is 0.17 V.[94] Hence, the OCV for Li/Na/K in bpn-BCN monolayer in the range of 0-1.8 V, particularly lower at higher concentrations, suggests that bpn-BCN is feasible as anode material for alkali metal ion batteries.

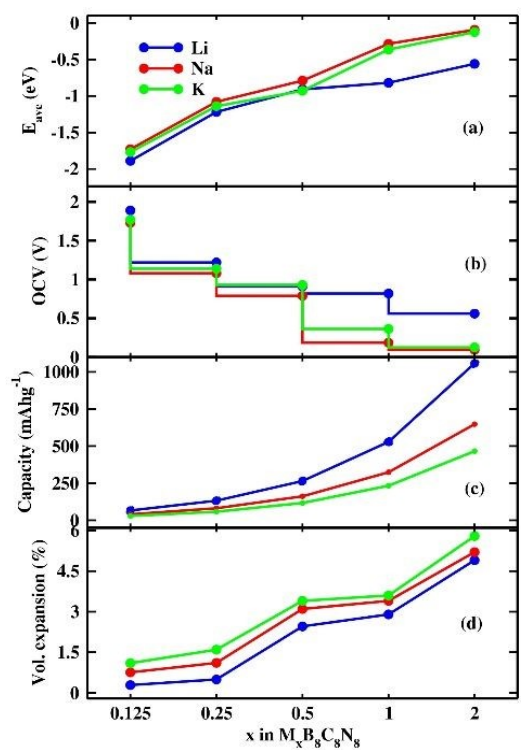

*Fig. 10* Shows the variation in (a) average adsorption energy, (b) open circuit voltage, (c) storage capacity, and (d) volume expansion of Li, Na, and K on bpn-BCN monolayers across different concentrations of metal atoms.

According to **eq. (10)**, the variation of storage capacity of bpn-BCN monolayer with alkali metal atoms concentration is shown in **Fig. 10(c).** The maximum calculated storage capacity of the bpn-BCN anode for Li, Na, and K is 1057.33, 647.27 and 465.98 mAhg$^{-1}$, respectively. The theoretical values of storage capacities for Li are three times greater than the commercially used graphite storage capacity.[91] The gravimetric capacity for Li is also higher than the other 2D dimensional materials such as graphene (744 mAhg$^{-1}$)[81], g-borocarbonitrides (710 mA h g$^{−1}$)[95], g-C$_3$N$_4$ (943 mAhg$^{-1}$)[96], T-graphene (876.55 mAhg$^{-1}$)[97] and comparable to carbon-based biphenylene monolayer (1075.37 mAhg$^{-1}$)[54] and BC$_2$N monolayer (1098 mAhg$^{-1}$).[98] Similarly, the storage capacity of bpn-BCN for Na and K is slightly lower than the previously reported carbon biphenylene monolayer.[99, 100]

To assess the impact of alkali metal atoms adsorption on the volume change of the bpn-BCN monolayer, we investigated the in-plane expansion of the bpn-BCN single-layer. As expected, the lattice parameter increases with increase M adsorption. Specifically, the in-plane lattice expansions for the maximum adsorption of Li, Na, and K are 4.90%, 5.20%, and 5.79%, respectively. The in-plane expansion of bpn-BCN with M concentration are shown in **Fig. 10 (d).** This trend aligns with the concept that larger atomic radii induce greater lattice expansion, evident in K having the largest radius among alkali metals, resulting in more volume expansion compared to Li/Na. Notably, the volume changes of the bpn-BCN monolayer during the adsorption/desorption of Li and non-Li atoms are comparatively smaller than those observed for graphite during the lithiation/delithiation process,[101] typically on the order of 10%. This indicates that expansion in the bpn-BCN monolayer during metal atoms adsorption is not a concern of worried.

We have examined energy barriers, storage capabilities, and open circuit voltage outcomes, which concluded that bpn-BCN shows promise as an active anode electrode in alkali ion batteries. The metal-adsorbed on bpn-BCN monolayer up to $x = 4$ have excellent structural stability with no structural distortion. High theoretical storage capacities and favourable OCV of bpn-BCN indicate it is a good electrochemical anode material for alkali ion batteries. However, ensuring the practical feasibility of this application strongly depends on the stability of the metal-adsorbed bpn-BCN monolayer at finite temperatures. To investigate the impact of temperature on alkali metal adsorbed monolayers, we performed ab-initio molecular dynamics (AIMD) simulations in the NVT canonical ensemble, maintaining a constant temperature of 500 K. A supercell of $3 \times 3 \times 1$ with a time step of 1 *fs* and a time range of 0-5 *ps* has been used in the framework of the NVT ensemble. **Fig. 11** shows the free energy vs. time plot and snapshots of each alkali metal-adsorbed monolayer at the end of 5 *ps* simulations.

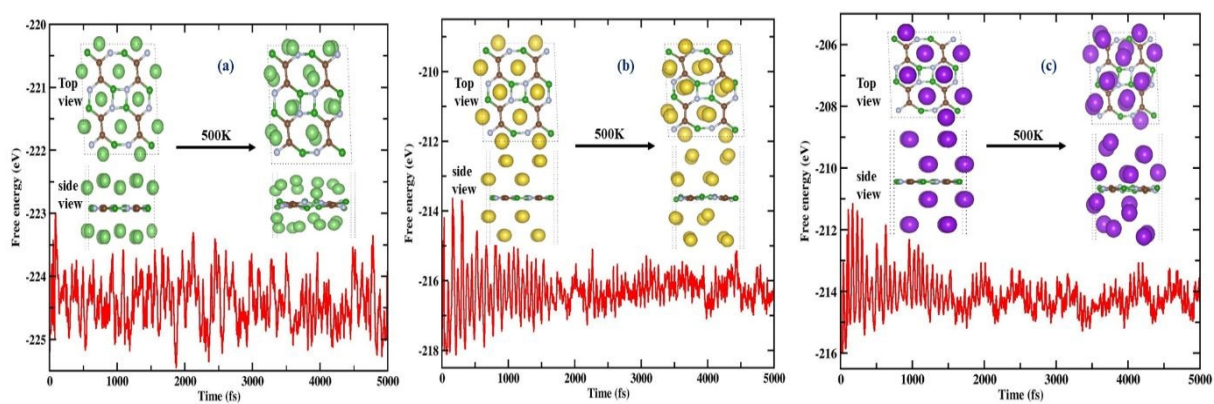

*Fig. 11* Presents the AIMD free energy versus time step plot, along with initial and final snapshots of the bpn-BCN biphenylene monolayer adsorbed with (a) Li, (b) Na, and (c) K. These simulations were conducted at 500 K over a 5 ps time scale.

The structural integrity of the bpn-BCN monolayer remains effectively preserved following metal adsorption during 5 *ps* of AIMD simulations. The AIMD-optimized geometry of all alkali metal atom-adsorbed bpn-BCN monolayers in **Fig. 11** shows that the bpn-BCN monolayer becomes slightly buckled, with a small increase in the interlayer distance of Li atoms. In contrast to Na and K intercalation, the interlayer distance between the alkali metal atoms has increased, as depicted in the snapshot attached in **Fig. 11**. The adsorbed alkali metal atoms deviate minimally from their energetically favourable sites. There are no appreciable deformations or bond breaks in the bpn-BCN monolayer during Li, Na, or K adsorption at 500K.

## Conclusions

In conclusion, hybrid borocarbonitrides phases, which are iso-electronic to experimentally synthesized carbon biphenylene 2D network BPN, have been explored by first-principles calculation. Six possible hybrid bpn-BCN phases iso-electronic to BPN with an equal stoichiometric ratio of B, C and N atoms have been studied. The phonon dispersion spectra with no negative frequency confirmed the dynamic stability of all phases. Further, the thermodynamics stability in terms of formation and cohesive energies revealed the free-standing feasibility of bpn-BCN phases. It has been revealed that the structures with maximum C-C, B-N and C-N bonds are more stable. Young's modulus and Poisson's ratio of all phases have been investigated for mechanical strength, comparable with the BPN. The AIMD calculation at 300K and 800K shows no major bonding distortion. Additionally, we explored the electrochemical properties of one of the most stable phases of the BCN-based biphenylene network as an anode material for metal (Li, Na, and K) ion batteries. Our calculations revealed that the BCN biphenylene monolayer can adsorb up to four layers of Li, Na, and K (two on each side), indicating large theoretical capacities for alkali-ion batteries. The calculated maximum storage capacities for Li, Na, and K are 1057.3, 647.27 and 465.98 mAhg$^{-1}$, respectively. The energy barrier for Li, Na and K atom migration between two hexagon centres via the square site is 0.65, 0.26 eV and 0.23 eV, respectively. These activation energies indicate quick diffusion of alkali atoms on the BCN biphenylene monolayer. Li-ion has a high diffusion barrier for BCN biphenylene monolayer compared to Na and K; it still has a high storage capacity. Furthermore, the average OCVs of BCN biphenylene monolayer for LiBs, NaBs, and KBs fall within a reasonable range of 0.34-1.30 V. Further, AIMD calculation reveals that the high symmetric H-, O- and S-sites are still energetically favourable at 500K. We conclude that the BCN biphenylene monolayer, with its enormous specific capacity, low energy barriers, outstanding structural stability, and moderate OCV values, holds great potential as a high-performance anode material for alkali metal batteries.

## Acknowledgement

AK thanks the University Grants Commission (UGC), New Delhi, for financial support through a Senior Research Fellowship (DEC18-512569-ACTIVE). PP thanks DST-SERB for the ECRA project (ECR/2017/003305).

## Data availability

The datasets generated during and/or analyzed during the current study are available from the corresponding author upon reasonable request.

## Code availability

Not applicable

## Conflict of interest

On behalf of all authors, the corresponding author states that there is no conflict of interest.